\documentclass[a4paper]{jpconf}

\usepackage{amsmath}
\usepackage{amssymb}
\usepackage{graphicx}
\usepackage{epsf}
\usepackage{slashed}
\usepackage{enumitem}
\usepackage[czech,english]{babel}
\usepackage[cp1250]{inputenc}
\usepackage{hyperref}
\usepackage{xcolor}
\usepackage{mathtools}

\usepackage[only,llbracket,rrbracket,llparenthesis,rrparenthesis]{stmaryrd} 
\usepackage{accsupp} 

\renewcommand{\e}{\ensuremath{\mathrm{e}}}
\newcommand{\eL}{\mathcal{L}}
\renewcommand{\d}{\ensuremath{\mathrm{d}}}
\newcommand{\threevector}[1]{\boldsymbol{#1}}
\newcommand{\qm}[1]{``#1''} 

\newcommand{\innerdot}{\ensuremath{\!\cdot\!}}

\newcommand{\group}[3]{#1(#2)_{\mathrm{#3}}}

\usepackage{bbm} 

\usepackage[nodayofweek]{datetime}
\newdateformat{mydate}{\twodigit{\THEDAY}{ }\shortmonthname[\THEMONTH], \THEYEAR}
\usepackage[normalem]{ulem}

\begin{document}

\title{Magnetic monopoles in extensions of Georgi--Glashow model}

\newcommand{\affiliacePraha}{Institute of Experimental and Applied Physics, \\ Czech Technical University in Prague, Husova~240/5, 110~00 Prague~1, Czech Republic}

\newcommand{\affiliaceOpava}{Research Centre for Theoretical Physics and Astrophysics, Institute of Physics, \\ Silesian University in Opava, Bezru\v{c}ovo n\'{a}m\v{e}st\'{\i}~1150/13, 746~01 Opava, Czech Republic}

\author{Petr Bene\v{s}$^1$ and Filip Blaschke$^{1,2}$}
\address{$^1$ \affiliacePraha}
\address{$^2$ \affiliaceOpava}

\ead{petr.benes@utef.cvut.cz}

\begin{abstract}
We discuss a class of effective extensions of the $\group{SU}{2}{}$ Georgi--Glashow model and discuss its Bogomol'nyi--Prasad--Sommerfield (BPS) limit. We identify a specific subclass of these models that admit analytical solutions of the monopole type. We present some concrete examples and find that the resulting monopoles tend to have their energy concentrated not in their center, but rather in a spherical shell around it.
\end{abstract}

\section{The model}

The simplest, if not prototypical model that supports the 't~Hooft--Polyakov magnetic monopole \cite{tHooft:1974kcl, Polyakov:1974ek} is the Georgi--Glashow model \cite{Georgi:1972cj}, i.e., spontaneously broken renormalizable $\group{SU}{2}{}$ gauge theory with adjoint (real triplet) scalar:
\begin{eqnarray}
\label{GeorgiGlashow}
\eL &=&
\frac{1}{2}
(D^\mu \threevector{\phi})^2
- \frac{1}{4g^2} (\threevector{F}^{\mu\nu})^2 
- V(\threevector{\phi}^2)
\,,
\end{eqnarray}
where
\begin{eqnarray}
D^\mu \threevector{\phi} &=& \partial^\mu \threevector{\phi} + \threevector{A}^\mu \times \threevector{\phi} \,,
\\
\threevector{F}^{\mu\nu} &=& \partial^\mu \threevector{A}^\nu - \partial^\nu \threevector{A}^\mu + \threevector{A}^\mu \times \threevector{A}^\nu \,.
\end{eqnarray}
The potential $V(\threevector{\phi}^2) \geq 0$ need not be specified, besides the key property that for $\threevector{\phi}^{2} = v^2$ (with $v^2 > 0$) it vanishes, triggering the spontaneous breakdown of the $\group{SU}{2}{}$ down to \qm{electromagnetic} $\group{U}{1}{}$ subgroup. The point is that embedding of this unbroken $\group{U}{1}{}$ in the parent $\group{SU}{2}{}$ can be different in different spatial directions, giving in turn rise to a topological defect -- the magnetic monopole.

The presence of the magnetic monopole in the spectrum of the Georgi--Glashow model \eqref{GeorgiGlashow} is given solely by the topological properties of the vacuum manifold, while the specific form of the interactions of the gauge bosons and scalars plays no r\^{o}le in this regard. However, the details of these interactions do influence the \qm{shape} of the monopoles, i.e., their energy density profiles.

Therefore, it was proposed in \cite{Benes:2023nsr} to release the assumption of renormalizability and study the monopole solutions in more general models. In particular, in order to constrain the vast space of models, it was proposed to consider the models whose Lagrangians are quadratic in both the derivatives of gauge fields, $\threevector{F}^{\mu\nu}$, and the derivatives of scalars, $D^\mu\threevector{\phi}$. Accordingly, the Lagrangian must be a linear combination of four algebraically independent terms: $(D^\mu \threevector{\phi})^2$ and $(\threevector{F}^{\mu\nu})^2$, that are present already in \eqref{GeorgiGlashow}, and the new terms $(\threevector{\phi} \innerdot D^\mu \threevector{\phi})^2$ and $(\threevector{\phi} \innerdot \threevector{F}^{\mu\nu} )^2$. Coefficients of this linear combination must be gauge-invariant functions of $\threevector{\phi}$. The most convenient way to write the corresponding Lagrangian turns out to be
\begin{eqnarray}
\label{lagrangian}
\eL &=&
\frac{v^2}{2} \bigg[
  f_1^2 \bigg( \frac{(D^\mu \threevector{\phi})^2}{\threevector{\phi}^2} - \frac{(\threevector{\phi} \innerdot D^\mu \threevector{\phi})^2}{\threevector{\phi}^4} \bigg)
+ f_3^2 \frac{(\threevector{\phi} \innerdot D^\mu \threevector{\phi})^2}{\threevector{\phi}^4}
\bigg]
\nonumber \\ && {}
- \frac{1}{4g^2} \bigg[
  f_2^2 \bigg( (\threevector{F}^{\mu\nu})^2 - \frac{(\threevector{\phi} \innerdot \threevector{F}^{\mu\nu})^2}{\threevector{\phi}^2} \bigg)
+ f_4^2 \frac{(\threevector{\phi} \innerdot \threevector{F}^{\mu\nu})^2}{\threevector{\phi}^2}
\bigg]
- V(\threevector{\phi}^2)
\,.
\end{eqnarray}
Here the coefficients $f_{i}^2$ are non-negative, dimensionless and gauge-invariant functions of $\threevector{\phi}$. We can make the last point more precise by noting that the scalar triplet can be decomposed as
\begin{eqnarray}
\label{definitionvHn}
\threevector{\phi} &=& v H \threevector{n} \,,
\end{eqnarray}
where the triplet $\threevector{n}$ is normalized as $\threevector{n}^2 = 1$.
Since the $\group{SU}{2}{}$ group acts only on $\threevector{n}$, the dimensionless form factor $H$ is gauge-invariant and we can conclude that
\begin{eqnarray}
f_i^2 &\equiv& f_i^2(H) \,.
\hspace{10mm}
(i=1,2,3,4)
\end{eqnarray}
Finally, we assume, without loss of generality, $f_{1,2}^2(1) = 1$ in order to have correct normalization of the kinetic terms in the vacuum.

We stress that \eqref{lagrangian} actually describes a whole class of models: Each set of the functions $f_i^2$ defines a specific theory. In particular, for $f_{1,3}^2 = H^2$ and $f_{2,4}^2 = 1$ we recover the Georgi--Glashow model \eqref{GeorgiGlashow}.

The classical Lagrange--Euler equations of motion
\begin{equation}
D_\mu \frac{\partial \eL}{\partial D_\mu\threevector{\phi}}
\ = \ 
\frac{\partial \eL}{\partial\threevector{\phi}}
\,,
\hspace{20mm}
D_\mu \frac{\partial \eL}{\partial \threevector{F}_{\mu\nu}} \ = \ 
\frac{1}{2} \threevector{\phi} \times \frac{\partial \eL}{\partial D_\nu\threevector{\phi}}
\end{equation}
recast for our Lagrangian \eqref{lagrangian} as
\begin{subequations}
\label{EOMsgeneral}
\begin{eqnarray}
D_\mu\bigg(f_1^2 \frac{v^2}{\threevector{\phi}^2} D^\mu\threevector{\phi}\bigg)
- \frac{\d}{\d H^2} \bigg(f_1^2\frac{v^2}{\threevector{\phi}^2}\bigg) (D^\mu\threevector{\phi})^2 \frac{\threevector{\phi}}{v^2}
&=& {}
\nonumber \\ && \hspace{-40mm} \ = \ 
- \frac{1}{2g^2} \bigg[\frac{\d f_2^{2}}{\d H^2} (\threevector{F}^{\mu\nu})^2 + \frac{\d (f_4^2-f_2^2)}{\d H^2} \frac{(\threevector{\phi}\innerdot\threevector{F}^{\mu\nu})^2}{\threevector{\phi}^2}\bigg] \frac{\threevector{\phi}}{v^2}
\nonumber \\ && \hspace{-40mm} \phantom{\ = \ }
- \frac{1}{2g^2} \frac{f_4^2-f_2^2}{\threevector{\phi}^4} \Big[\threevector{\phi}^2 (\threevector{\phi}\innerdot\threevector{F}_{\mu\nu}) \threevector{F}^{\mu\nu} - (\threevector{\phi}\innerdot\threevector{F}^{\mu\nu})^2 \threevector{\phi}\Big]
\nonumber \\ && \hspace{-40mm} \phantom{\ = \ }
- \bigg[
\frac{\d (f_3^2-f_1^2)}{\d H^2} \frac{(\partial^\mu\threevector{\phi}^2)^2}{4\threevector{\phi}^4}
+ \frac{v^2}{2\threevector{\phi}^2} (f_3^2-f_1^2) \bigg(\frac{\partial_\nu \partial^\nu\threevector{\phi}^2}{\threevector{\phi}^2}
- \frac{(\partial^\mu\threevector{\phi}^2)^2}{\threevector{\phi}^4}\bigg)
\bigg] \threevector{\phi}
\nonumber \\ && \hspace{-40mm} \phantom{\ = \ }
- 2 \frac{\d V}{\d H^2} \frac{\threevector{\phi}}{v^2}
\,,
\\
\frac{1}{g^2} D_\nu \bigg[f_2^2 \threevector{F}^{\nu\mu} + (f_4^2-f_2^2) \frac{(\threevector{\phi}\innerdot\threevector{F}^{\nu\mu})}{\threevector{\phi}^2} \threevector{\phi}\bigg] &=& f_1^2 \frac{v^2}{\threevector{\phi}^2} (D^\mu\threevector{\phi}) \times \threevector{\phi}
\,.
\end{eqnarray}
\end{subequations}
Alas, these equations are way too complicated to be solved analytically. Luckily, there is a trick due to Bogomol'nyi, Prasad and Sommerfield \cite{Bogomolny:1975de,Prasad:1975kr} that allows to simplify these equations and reduce their order. If the functions $f_i^2$ satisfy the key condition
\begin{eqnarray}
\label{relationBPSfi}
f_3 f_4 &=& H \frac{\d}{\d H} \big(f_1 f_2 \big) \,,
\end{eqnarray}
then
the energy density of a static configuration $\partial^0 = \threevector{A}^0 = 0$ (in which we are interested in) can be written as
\begin{eqnarray}
\label{Edens}
\mathcal{E} &=&
\frac{1}{2} \bigg[
  \frac{f_1}{H} D^i \threevector{\phi}
- \frac{f_2}{g} \threevector{B}^i
+ \bigg(
  \frac{(f_2+f_4)(\threevector{\phi} \innerdot \threevector{B}^i)}{g}
- \frac{(f_1+f_3)(\threevector{\phi} \innerdot D^i \threevector{\phi})}{H}
\bigg) \frac{\threevector{\phi}}{\threevector{\phi}^2}
\bigg]^2
+ \partial^i \bigg(\frac{f_1 f_2}{gH} \threevector{\phi} \innerdot \threevector{B}^i\bigg)
\nonumber \\ && {}
+ V
\,,
\end{eqnarray}
where $\threevector{B}^i \equiv -\frac{1}{2} \epsilon^{ijk} \threevector{F}^{jk}$. Let us now consider the Bogomol'nyi--Prasad--Sommerfield (BPS) limit, i.e., the limit of vanishing potential, $V \to 0$, but with the boundary condition $\threevector{\phi}^2 = v^2$ kept. Now, if the square in \eqref{Edens} vanishes, the energy density is a total derivative, so that it is determined solely by the boundary conditions and not by the equations of motion. Since the boundary conditions are fixed, the energy density is then minimized and the so-called Bogomol'nyi energy bound is saturated. Necessarily, the corresponding field configuration must be a solution of the equations of motion. Therefore, in the BPS limit the condition for the square in \eqref{Edens} to vanish is equivalent to the full equations of motion \eqref{EOMsgeneral}. This condition is called the BPS equation of motion and can be, after some rearrangements, written as
\begin{eqnarray}
\label{BPSeqnFi}
D^i \threevector{\phi}
&=&
\frac{H}{g} \bigg[
\frac{f_2}{f_1} \bigg(\threevector{B}^i
- \frac{\threevector{\phi} \innerdot \threevector{B}^i}{\threevector{\phi}^2}\threevector{\phi}\bigg)
+ \frac{f_4}{f_3} \frac{\threevector{\phi} \innerdot \threevector{B}^i}{\threevector{\phi}^2} \threevector{\phi}
\bigg]
\,.
\end{eqnarray}
Notice that the BPS equation is much simpler than \eqref{EOMsgeneral}, not to mention that it is only of the first and not the second order. This gives some hope that it could be, at least in some cases, analytically solvable.


Let us now adopt the spherically symmetric \qm{hedgehog} Ansatz
\begin{equation}
\label{ansatz}
\threevector{\phi} \ = \ v H \frac{\threevector{r}}{r} \,,
\hspace{10mm}
\threevector{A}^i \ = \
- \frac{\threevector{r} \times \partial^i \threevector{r}}{r^2} (1-K)
\,,
\end{equation}
where the form factors $H$ and $K$ are functions of $r = |\threevector{r}|$ and satisfy the  boundary conditions $H(\infty) = 1$ and $K(\infty) = 0$
that follow from the requirement that the total energy be finite. The BPS equation \eqref{BPSeqnFi} under the Ansatz \eqref{ansatz} decouples into a system of two ordinary differential equations for $K$ and $H$:
\begin{equation}
\label{logKH}
\partial_\rho(\log K) \ = \ - \frac{f_1}{f_2}
\,,
\hspace{10mm}
\partial_\rho (\log H) \ = \ 
\frac{1-K^2}{\rho^2}
\frac{f_4}{f_3}
\,,
\end{equation}
where we introduced a  dimensionless radius
\begin{eqnarray}
\rho &\equiv& vgr \,.
\end{eqnarray}
The energy density reads 
\begin{eqnarray}
\frac{\mathcal{E}}{v^4g^2} &=&
\frac{1}{\rho^2} \partial_\rho \Big[f_1 f_2 \big(1-K^2\big) \Big]
\,,
\end{eqnarray}
so that the mass of the monopole is
\begin{eqnarray}
M &=& \int_{\mathbb{R}^3}\!\d^3 x \, \mathcal{E}
\ = \ \frac{4 \pi v}{g} \int_{0}^{\infty}\!\d\rho \, \rho^2 \frac{\mathcal{E}}{v^4 g^2}
\ = \ 
\frac{4 \pi v}{g}
\,.
\end{eqnarray}
Finally, the magnetic charge of the monopole comes out to be $q_{\mathrm{m}} = 4\pi$. Notice that both mass and magnetic charge of the monopole are independent of the functions $f_i^2$.

The equations \eqref{logKH} are simple enough to be analytically solvable in some cases. One such case, studied in detail in \cite{Benes:2023nsr}, is when
\begin{eqnarray}
\label{condF4}
\frac{f_3}{f_4} &=& H \frac{\d}{\d H}\bigg(\frac{f_1}{f_2}\bigg) \,.
\end{eqnarray}
(Notice similarity with the BPS condition \eqref{relationBPSfi} that has to be, of course, satisfied as well.) The solution to \eqref{logKH} is then simply
\begin{equation}
\label{solutiongeneral}
K \ = \ \frac{\rho}{\sinh \rho}
\,,
\hspace{10mm}
H \ = \ 
\bigg(\frac{f_1}{f_2}\bigg)^{-1} \bigg(\coth\rho-\frac{1}{\rho}\bigg)
\,.
\end{equation}

\section{Examples}

\subsection{Power function}

As the simplest example, let's consider
\begin{subequations}
\begin{align}
f_1^2 &\ = \ 
H^{n+m} \,,
&
f_3^2 &\ = \ 
nm f_1^2 \,,
\\
f_2^2 &\ = \ 
H^{n-m} \,,
&
f_4^2 &\ = \ 
\frac{n}{m} f_2^2 \,.
\end{align}
\end{subequations}
where $n \geq m$ are parameters of the model. This is the most general model with all $f_i^2$ being power functions that satisfy both \eqref{relationBPSfi} and \eqref{condF4}. The corresponding Lagrangian reads
\begin{eqnarray}
\label{lagrangianPowerInv}
\eL &=&
\frac{v^2}{2} H^{n+m} \bigg[
\frac{(D^\mu \threevector{\phi})^2}{\threevector{\phi}^2}
+ (nm-1) \frac{(\threevector{\phi} \innerdot D^\mu \threevector{\phi})^2}{\threevector{\phi}^4}
\bigg]
- \frac{1}{4g^2} H^{n-m} \bigg[
(\threevector{F}^{\mu\nu})^2
+ \bigg(\frac{n}{m}-1\bigg) \frac{(\threevector{\phi} \innerdot \threevector{F}^{\mu\nu} )^2}{\threevector{\phi}^2}
\bigg]
\,.
\nonumber \\ && {}
\end{eqnarray}
For $n = m = 1$ this is nothing but the Georgi--Glashow model \eqref{GeorgiGlashow}. Using \eqref{solutiongeneral} we readily find the solution to be $K = \rho / \sinh \rho$ and $H = (\coth\rho - 1 / \rho)^{1/n}$.

Let us now focus on the energy density. It is of course a total derivative, but using the BPS equation it can be written also in the form
\begin{equation}
\label{EdensPower}
\frac{\mathcal{E}}{v^4g^2}
\ = \ 
\kappa^{\frac{n}{m}} \bigg[2 \kappa \frac{K^2}{\rho^2}
+ \frac{n}{m} \frac{1}{\kappa} \frac{(1-K^2)^2}{\rho^4}\bigg]
\,,
\end{equation}
where we denoted $\kappa = \coth\rho - 1 / \rho$. First of all, notice that while the Lagrangian \eqref{lagrangianPowerInv} depends on the parameters $n$ and $m$ \qm{independently}, the physical, measurable quantity $\mathcal{E}$ depends only on their particular combination $n/m$. This is not a coincidence. As thoroughly discussed in \cite{Benes:2023nsr}, it is a consequence of the \emph{form-invariance} of our general class of theories \eqref{lagrangian} under redefinitions of $H$. In this particular case \eqref{lagrangianPowerInv}, theories with different $n$ and $m$, but with the same $n/m$ can be shown to be physically equivalent, as exemplified in \eqref{EdensPower}.

\begin{figure}[t]
\begin{center}
\includegraphics[width=0.8\textwidth]{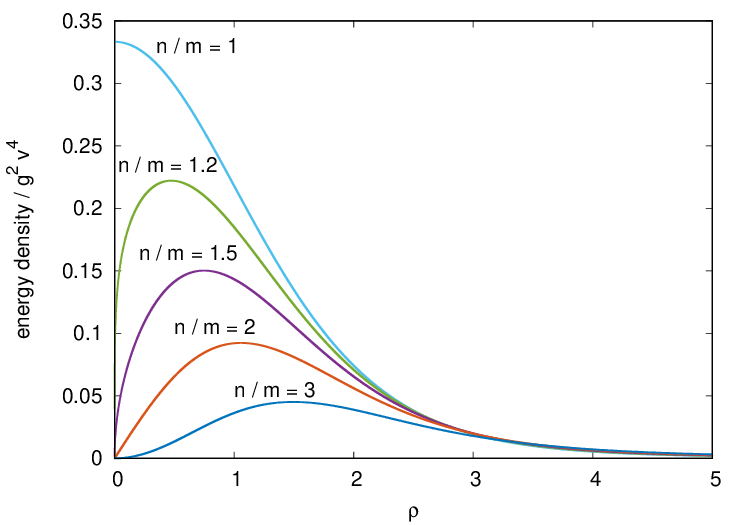}
\caption{Energy density for a single monopole solution of the power-function theory \eqref{lagrangianPowerInv} for various values of $n/m \geq 1$.}
\label{plot_inv_pow_E}
\end{center}
\end{figure}

In Fig.~\ref{plot_inv_pow_E} we plot the energy density for several values of $n/m$. There are two types of behavior. For $n/m = 1$ we obtain the energy density that is concentrated in the monopole core. This is a sort of usual and expected behavior, as it is what we are accustomed to from the \qm{canonical} 't~Hooft--Polyakov monopole in the Georgi--Glashow model. (After all, the case $n/m = 1$ actually \emph{is} the Georgi--Glashow model.) However, for $n/m > 1$ we see something different: The energy density vanishes in the center of the monopole and is peaked in a finite distance from it; in other words, most of the monopole energy (mass) is stored in a spherical shell around its center. Hence this type of solutions can be dubbed \emph{hollow monopoles} \cite{Bazeia:2018fhg}.

\subsection{Power-exponential function}

The previous model can be generalized as
\begin{subequations}
\begin{align}
f_1^2 &\ = \ 
H^{n+m} \e^{a(H^{n\ell}-1)^k} \,,
&
f_3^2 &\ = \ 
nm \Big[1 + a\ell k\big(H^{n\ell}-1\big)^{k-1} H^{n\ell}\Big] f_1^2 \,,
\\
f_2^2 &\ = \ 
H^{n-m} \e^{a(H^{n\ell}-1)^k} \,,
&
f_4^2 &\ = \ 
\frac{n}{m} \Big[1 + a\ell k\big(H^{n\ell}-1\big)^{k-1} H^{n\ell}\Big] f_2^2 \,,
\end{align}
\end{subequations}
that is,
\begin{eqnarray}
\label{LagPowExp}
\eL &=&
\hspace{6mm}
\frac{v^2}{2} H^{n+m} \bigg[
\frac{(D^\mu \threevector{\phi})^2}{\threevector{\phi}^2}
+ \Big(nm \Big[1 + a\ell k\big(H^{n\ell}-1\big)^{k-1} H^{n\ell}\Big] -1\Big) \frac{(\threevector{\phi} \innerdot D^\mu \threevector{\phi})^2}{\threevector{\phi}^4}
\bigg] \e^{a(H^{n\ell}-1)^k}
\nonumber \\ && {}
- \frac{1}{4g^2} H^{n-m} \bigg[
(\threevector{F}^{\mu\nu})^2
+ \bigg(\frac{n}{m} \Big[1 + a\ell k\big(H^{n\ell}-1\big)^{k-1} H^{n\ell}\Big]-1\bigg) \frac{(\threevector{\phi} \innerdot \threevector{F}^{\mu\nu})^2}{\threevector{\phi}^2}
\bigg] \e^{a(H^{n\ell}-1)^k}
\,.
\hspace{8mm}
\end{eqnarray}
Now there are five parameters: $a$, $k$, $\ell$, $m$, $n$.

\begin{figure}[t]
\begin{center}
\includegraphics[width=0.8\textwidth]{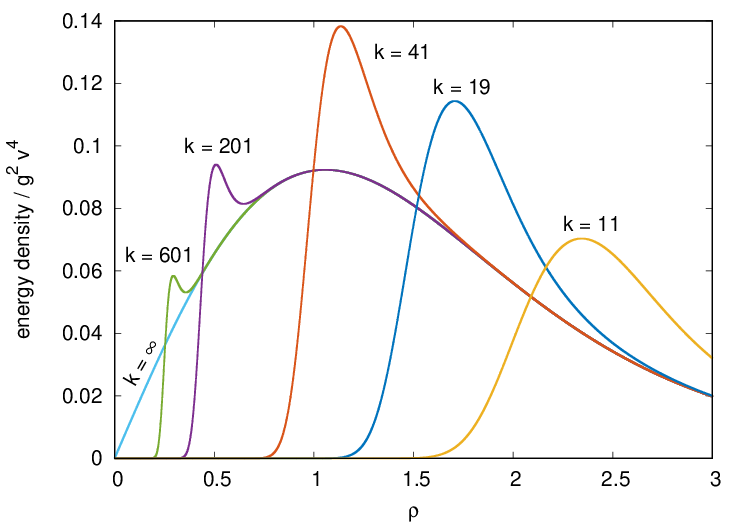}
\caption{Energy density for a single monopole solution of the theory \eqref{LagPowExp} with fixed $n/m = 2$, $\ell = 1$, $a = 100$ and various values of $k$.}
\label{plot_inv_powexp_2min_1_E}
\end{center}
\end{figure}

\begin{figure}[t]
\begin{center}
\includegraphics[width=0.8\textwidth]{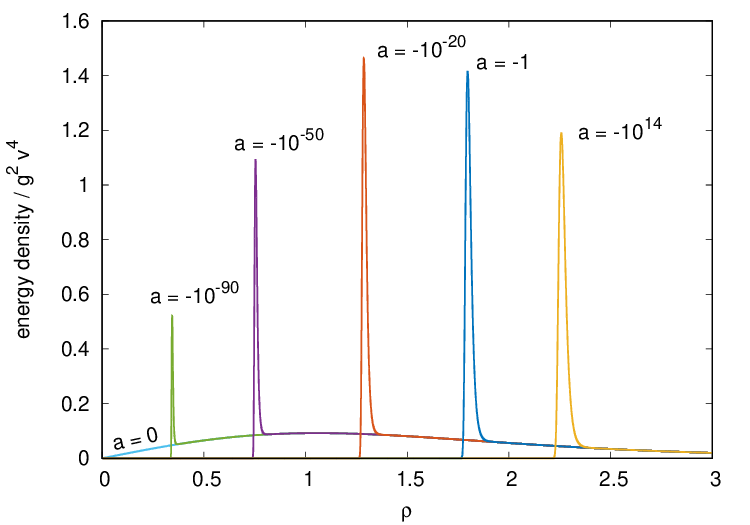}
\caption{Energy density for a single monopole solution of the theory \eqref{LagPowExp} with fixed $n/m = 2$, $\ell = -1/2$, $k = 101$ and various values of $a$.}
\label{plot_inv_powexp_1min_exp_E}
\end{center}
\end{figure}

The \emph{raison d'\^{e}tre} of this rather complicated and somewhat artificial model is to showcase even more interesting energy density profiles than in the previous power-function model. In Fig.~\ref{plot_inv_powexp_2min_1_E} we show the energy density for various values of $k$, with the other parameters fixed as $n/m = 2$, $\ell = 1$, $a = 100$. While we again obtain hollow monopoles, this time the energy density can have not just one, but two maxima, so that the shell of energy around the monopole center can be structured, with two \qm{sub-shells}. In Fig.~\ref{plot_inv_powexp_1min_exp_E} we see that the peak of the energy density of a hollow monopoles can be much sharper than in the previous examples. Moreover, while not visible from the plot, the energy densities in Fig.~\ref{plot_inv_powexp_1min_exp_E} actually fall off exponentially as $\rho \to 0$. Accordingly, there is a finite region in the center of the monopoles of this type with virtually vanishing energy density.

\section{Conclusions}

We discussed a particular class of effective extensions \eqref{lagrangian} of the Georgi--Glashow model \eqref{GeorgiGlashow}. We found that in order to obtain equation of motion of the first order in the limit of vanishing potential (a.k.a.~the BPS limit), the functions defining the model have to be related to one another as \eqref{relationBPSfi}. Moreover, if these functions satisfy also relation \eqref{condF4}, it is possible to find single monopole solutions in a closed analytic form. We presented several explicit examples of these analytic solutions. We found that the resulting monopoles typically have their energy density concentrated not in the very center of the monopole, but rather in a spherical shell around it. Moreover, the shell itself can be structured, with several sub-shells.

\ack

The authors would like to express the gratitude for the institutional support of the Institute of Experimental and Applied Physics, Czech Technical University in Prague (P.~B.~and F.~B.), and of the Research Centre for Theoretical Physics and Astrophysics, Institute of Physics, Silesian University in Opava (F.~B.). P.~B.~is indebted to A\v{s}tar \v{S}eran and FSM for invaluable discussions.

\section*{References}


\begin{thebibliography}{1}
\expandafter\ifx\csname url\endcsname\relax
  \def\url#1{{\tt #1}}\fi
\expandafter\ifx\csname urlprefix\endcsname\relax\def\urlprefix{URL }\fi
\providecommand{\eprint}[2][]{\url{#2}}

\bibitem{tHooft:1974kcl}
't~Hooft G 1974 {\em Nucl. Phys. B\/} {\bf 79} 276--284

\bibitem{Polyakov:1974ek}
Polyakov A~M 1974 {\em JETP Lett.\/} {\bf 20} 194--195

\bibitem{Georgi:1972cj}
Georgi H and Glashow S~L 1972 {\em Phys. Rev. Lett.\/} {\bf 28} 1494

\bibitem{Benes:2023nsr}
Bene\v{s} P and Blaschke F 2023 {\em Phys. Rev. D\/} {\bf 107} 125002
  (\textit{Preprint} \eprint{2303.15602})

\bibitem{Bogomolny:1975de}
Bogomolny E~B 1976 {\em Sov. J. Nucl. Phys.\/} {\bf 24} 449

\bibitem{Prasad:1975kr}
Prasad M~K and Sommerfield C~M 1975 {\em Phys. Rev. Lett.\/} {\bf 35} 760--762

\bibitem{Bazeia:2018fhg}
Bazeia D, Marques M~A and Olmo G~J 2018 {\em Phys. Rev. D\/} {\bf 98} 025017
  (\textit{Preprint} \eprint{1807.01299})

\end{thebibliography}

\providecommand{\newblock}{}

\end{document}